\documentclass[journal,draftcls,onecolumn,12pt,twoside]{IEEEtranTCOM}

\usepackage{amsmath,amssymb,epsfig,psfrag,cite,subfigure}
\usepackage{dblfloatfix}    
\usepackage{graphicx}
\usepackage[font=footnotesize]{caption}
\usepackage{algorithm}
\usepackage{epsfig,psfrag}
\usepackage{subfigure}
\usepackage{color}
\usepackage{url}
\usepackage[margin=1 in]{geometry}

\usepackage{algorithm}
\usepackage{algpseudocode}
\usepackage{placeins}

\newcommand{\NL}{{N_{\rm{L}}}}
\newcommand{\NR}{{N_{\rm{I}}}}

\newcommand{\mi}{{\rm{m}}_i}

\newcommand{\x}{\boldsymbol{x}}

\newcommand{\PTXi}{P_{{\rm{TX}},i}}
\newcommand{\PTXj}{P_{{\rm{TX}},j}}

\newcommand{\PRXi}{P_{{\rm{RX}},i}}
\newcommand{\PRXj}{P_{{\rm{RX}},j}}

\newcommand{\PRXone}{P_{{\rm{RX}},1}}
\newcommand{\PRXlast}{P_{{\rm{RX}},\NL}}
\newcommand{\PRX}{{\boldsymbol{P}}_{{\rm{RX}}}}
\newcommand{\nT}{{\boldsymbol{n}}_{{\rm{T}}}}
\newcommand{\nR}{{\boldsymbol{n}}_{{\rm{R}}}}

\newcommand{\mtL}{{\mathcal{L}}}

\newcommand{\Ar}{{A_{\mathrm{R}}}}
\newcommand{\norm}[1]{\big\lVert#1\big\rVert}

\newcommand{\tl}{\boldsymbol{\tilde{l}}}
\newcommand{\tn}{\boldsymbol{\tilde{n}}}
\newcommand{\lt}{\boldsymbol{{l}}}
\newcommand{\Ni}{\boldsymbol{{n}_i}}
\newcommand{\xhatmml}{{\boldsymbol{\hat{x}}}_{{\rm{MML}}}}
\newcommand{\xhat}{{\boldsymbol{\hat{x}}}}
\newcommand{\xbar}{\bar{{\boldsymbol{x}}}}
\renewcommand{\baselinestretch}{1.8}

\begin{document}

\include{macros}

\title{Visible Light Positioning with Intelligent Reflecting Surfaces under Mismatched Orientations}

\author{Issifu Iddrisu and\thanks{I. Iddrisu and S. Gezici are with the Department of Electrical and Electronics Engineering, Bilkent University, Ankara 06800, Turkey (e-mails: issifu.iddrisu@bilkent.edu.tr,\,gezici@ee.bilkent.edu.tr).} Sinan Gezici}

\maketitle

\vspace{-2cm}

\begin{abstract}
Accurate localization can be performed in visible light systems in non-line-of-sight (NLOS) scenarios by utilizing intelligent reflecting surfaces (IRSs), which are commonly in the form of mirror arrays with adjustable orientations. When signals transmitted from light emitting diodes (LEDs) are reflected from IRSs and collected by a receiver, the position of the receiver can be estimated based on power measurements by utilizing the known parameters of the LEDs and IRSs. Since the orientation vectors of IRS elements (mirrors) cannot be adjusted perfectly in practice, it is important to evaluate the effects of mismatches between desired and true orientations of IRS elements. In this study, we derive the misspecified Cramér-Rao lower bound (MCRB) and the mismatched maximum likelihood (MML) estimator for specifying the estimation performance and the lower bound in the presence of mismatches in IRS orientations. We also provide comparisons with the conventional maximum likelihood (ML) estimator and the CRB in absence of orientation mismatches for quantifying the effects of mismatches. It is shown that orientation mismatches can result in significant degradation in localization accuracy at high signal-to-noise ratios.

\end{abstract}

\section{Introduction}\label{sec:Intro}

In recent research studies, visible light positioning (VLP) systems have attracted significant attention due to their cost-effectiveness and high accuracy compared to radio frequency (RF)-based positioning systems \cite{SurveyVLPprocIEEE,PIEEE,Yazar_vlp2020}. By utilizing the widespread availability of light-emitting diodes (LEDs) as indoor lighting sources, these systems offer minimal deployment costs. VLP systems are particularly suitable for indoor applications that require high localization accuracy such as navigational services for both humans and robots, and asset tracking \cite{roadmap,Survey_VLP,Monica_UWB,EnLighting,Dimian2017,TDOA_VLP_Automative,Platoon_VLP}. In the literature, numerous position estimation algorithms have been developed, and comprehensive analyses of theoretical accuracy limits for VLP systems have been performed. The received signal strength (RSS) parameter is widely utilized in VLP systems due to its cost-effective hardware implementation and high accuracy \cite{PIEEE}. In \cite{VLP_CRLB_RSS}, the Cramér–Rao bound (CRB) is derived as the theoretical accuracy limit of RSS based VLP systems. Also, in \cite{onTheFundamental}, the authors derived closed-form CRB expressions for the location and orientation of a visible light communication (VLC) receiver, taking into account the effects of non-line-of-sight (NLOS) propagation on VLP performance.

More recently, the emergence of intelligent reflecting surface (IRS), also referred to as reconfigurable intelligent surface (RIS), has drawn significant attention due to its considerable advantages in widening wireless network coverage (specifically for RF systems), enhancing wireless communication rates, and reducing deployment and energy costs \cite{basar2019wireless,liaskos2018new}. An IRS is a planar surface made up of numerous low-cost reflecting elements, each capable of independently altering the amplitude and/or phase of an incident signal \cite{Wu_IRS}. In \cite{wu2019intelligent}, the authors propose an IRS-based method to improve spectrum and energy efficiency while reducing the implementation cost of wireless communication systems. Also, the authors of \cite{yu2019miso} investigate the joint design of the beamformer at the transmitter and the IRS phase shifts for an IRS-assisted wireless communication system, and demonstrate that IRSs hold significant potential for creating high-speed and energy-efficient communication networks. 

There are also a number of studies in the literature that explore IRS-assisted VLC systems. Conventionally, communication performance of IRS systems heavily depends on the presence of line-of-sight (LoS) paths. IRS, capable of re-configuring the wireless propagation channel, mitigates the LoS requirement in VLC systems. In \cite{abdelhady2020visible}, two intelligent reflection systems using programmable metasurfaces and mirrors to concentrate incident light onto the VLC receiver are proposed, and the detector plane irradiance expressions for both systems to evaluate their aiming and focusing performance are derived. The authors in \cite{aboagye2021intelligent} investigate a VLC system incorporating an IRS mirror array to address LoS blockage issues caused by the random orientation of the user device, self-blockage, and external obstructions. In \cite{eldeeb2023energy}, the authors employ optical IRSs to relax the LoS constraint in vehicle-to-vehicle (V2V) VLC systems and derive a closed-form expression for the number of IRS elements needed to achieve a desired energy or spectral efficiency. The authors in \cite{wu2022configuring} derive an IRS-aided parallel multiple-input-multiple-output (MIMO) VLC channel model and demonstrate that incorporating IRS into such a system significantly enhances its asymptotic capacity. Also, in \cite{sun2023optical}, the authors derive asymptotic capacity expressions for optical IRS-assisted MIMO VLC under various emission power constraints in the high-SNR regime and propose two algorithms to maximize these capacities by jointly optimizing the alignment of optical IRS elements and the transmitted power. In \cite{xu2024energy}, the authors investigate a downlink energy efficiency maximization problem in an IRS-VLC system by formulating a resource allocation problem based on achievable system sum rate versus total power consumption, under user data and transmit power constraints.

While a significant amount of research is focused on IRS-assisted RF communications, there are also a number of studies in the literature that examine the utilization of IRSs in RF-based positioning systems \cite{wymeersch2020beyond,zhou2022aoa,alexandropoulos2022localization,hua2023intelligent,zhang2020towards}.
In \cite{zhou2022aoa}, the authors investigate the utilization of IRSs for user positioning and introduce an angle-domain method to acquire angle of arrival (AoA) data for the direct path between users and IRSs, subsequently employing AoA positioning at the access point (AP) for user localization. In \cite{hua2023intelligent}, the authors investigate the target localization problem utilizing semi-passive IRS and apply the MUSIC algorithm to estimate the target direction.
Also, in \cite{alexandropoulos2022localization}, a maximum likelihood (ML)-based localization method utilizing multiple single-receive RF IRSs, which employs beamspace orthogonal matching pursuit (OMP) for AoA estimation and least squares line intersection is presented.
In \cite{wymeersch2020beyond}, the authors investigate a downlink positioning problem aided by IRSs, analyze it through Fisher information and propose a two-step scheme to optimize IRSs to improve positioning performance.
Additionally, in \cite{zhang2020towards}, an IRS-aided multi-user positioning protocol is proposed, and the optimization problem for positioning is formulated and solved using a configuration optimization algorithm. 

To the best of our knowledge, there exist only two studies in the literature on IRS-assisted visible light positioning (VLP) \cite{wang2023ris ,kokdogan2024intelligent}. The authors in \cite{wang2023ris} propose an algorithm based on sparse Bayesian learning for solving the localization problem in IRS-assisted indoor VLP, deriving the Fisher information matrix, position estimation bound (PEB), and RMSE expressions for NLOS scenarios. In \cite{kokdogan2024intelligent}, the authors formulate an RSS-based position estimation problem in a VLP system utilizing IRSs. An ML estimator is derived utilizing signals from both the LoS and IRS elements, along with the corresponding CRB, and an algorithm is proposed to optimally adjust the IRS elements to maximize the received power. However, this study assumes a perfect alignment between the desired orientations of IRS elements and their actual orientations, an ideal scenario rarely achieved in practical settings due to inaccuracies in gyroscope readings and noise.

In this manuscript, we investigate the problem of IRS-assisted VLP, particularly focusing on a scenario where there is a mismatch between the actual orientations of the IRS elements and the desired orientations assumed by the VLC receiver. To this aim, we employ a misspecified Cramér-Rao bound (MCRB) analysis \cite{fortunati2017} to quantify the deterioration in positioning performance due to the mismatch between the true orientations of the IRS elements and the assumed orientations at the receiver. The main contributions and novel aspects of this study can be summarized as follows:
\begin{itemize}
\item The problem of IRS-assisted VLP under mismatched orientations is formulated for the first time in the literature.
\item A mismatched maximum likelihood (MML) estimator is derived and the corresponding MCRB and lower bound (LB) expressions are obtained, serving as benchmarks for evaluating localization performance.
\item A comparative analysis is performed between the MML estimator and the maximum likelihood (ML) estimator (and also between the MCRB and the CRB) to reveal the effects of orientation mismatches on localization performance.
\end{itemize}

The remainder of the manuscript is organized as follows: Section~\ref{sec:SysModel} describes the VLP system and the signal and channel models. Section~\ref{sec:Est} focuses on the theoretical results by deriving the MML estimator and the MCRB in the presence of orientation mismatches. The comparative numerical results are provided in Section~\ref{sec:Nume} by presenting the estimators and the bounds in the presence and absence of orientation mismatches. Finally, concluding remarks are made in Section~\ref{sec:Conc}.


\section{System Model}\label{sec:SysModel}

Consider a VLP system with $\NL$ LED transmitters at known locations denoted by $\lt_i\in{\mathbb{R}}^3$ for $i\in\{1,\ldots,\NL\}$ and a VLC receiver at an unknown location $\x\in{\mathbb{R}}^3$. The orientation vectors of the $i$th LED transmitter and the VLC receiver are denoted by $\Ni\in{\mathbb{R}}^3$ and $\nR\in{\mathbb{R}}^3$, respectively. Also, there is an IRS with $\NR$ elements (small mirrors) denoted by $S_1,\ldots,S_\NR$ having orientation vectors ${\boldsymbol{\tilde{n}}}_1,\ldots,{\boldsymbol{\tilde{n}}}_\NR$ and locations ${\boldsymbol{\tilde{l}}}_1,\ldots,{\boldsymbol{\tilde{l}}}_\NR$. It is assumed that each IRS element causes glossy reflections (Phong’s model) \cite{GlossyTCOM,abdelhady2020visible} with the reflectance coefficient remaining constant across each element, where $\rho_k$ denotes the reflectance coefficient for the $k$th IRS element. It is assumed that both $\rho_k$'s and $S_k$'s are assumed to be known for $k \in \{1,2,\ldots,\NR\}$ \cite{kokdogan2024intelligent}.

The VLC receiver aims to determine its own position using signals received from the LED transmitters. To this aim, it gathers power measurement from the LED transmitters. The power measurement received by the VLC receiver due to the signal emitted by the $i$th LED transmitter is expressed as \cite{kokdogan2024intelligent}
\begin{gather}
   \PRXi = P_{{\rm{TX}},i} H^{\rm{LOS}}_{i}(\x) +P_{{\rm{TX}},i} \sum_{k=1}^{\NR} \int_{S_{k}} dH_{i,k}^{{\rm{ref}}} (\x,\tl_k,\tn_k) + \eta_i
   \label{eq:powMeas}
\end{gather}
for $i=1,\ldots,\NL$, where $\PRXi$ is the received power at the VLC receiver due to the signal from the $i$th LED transmitter, $P_{{\rm{TX}},i}$ is the transmit power of the $i$th LED transmitter, $H^{{\rm{LOS}}}_{i}(\x)$ denotes the line-of-sight (LOS) channel coefficient between the $i$th LED transmitter and the VLC receiver, $dH_{i,k}^{{\rm{ref}}}(\x,\tl_k,\tn_k)$ denotes the reflected channel coefficient of the path between the $i$th LED transmitter and the VLC receiver, which makes a single reflection from an infinitesimally small area around $\tl_k$ at the $k$th IRS element, and $\eta_i$ is zero-mean Gaussian noise with a variance of $\sigma_i^2$, which is independent of $\eta_j$ for all $j\ne i$. As indicated in \eqref{eq:powMeas}, it is presumed that the signals emitted by different LED transmitters undergo separate processing, and their individual power levels are measured at the VLC receiver. This assumption relies on employing a specific form of multiple access protocol, such as time-division or frequency-division multiple access \cite{VLP_power_allocation}.

The LOS channel gain in \eqref{eq:powMeas} can be calculated as 
\begin{gather}
H^{{\rm{LOS}}}_{i}(\x) = \frac{(\mi+1) \Ar\left[(\x - \lt_i)^{T} \Ni \right]^{\mi} (\lt_i-\x)^{T} \nR}{2\pi\norm{\x - \lt_i}^{\mi+3}} 
\label{eq:directChannel}
\end{gather}
and the reflected channel gain in \eqref{eq:powMeas} can be expressed as follows \cite{kokdogan2024intelligent}:
\begin{align}\nonumber 
dH_{i,k}^{{\rm{ref}}} (\x,\tl_k,\tn_k) =& \frac{(\mi+1) \left[(\tl_k - \lt_i)^{T} \Ni \right]^{\mi} ((\lt_i-\tl_k)^{T} \tn_k)}{4\pi^2\norm{\lt_i - \tl_k}^{\mi+3}\norm{\x - \tl_k}^{3}} (\Ar\rho_k((\tl_k-\x)^{T} \nR)dS_k\\&\times\left(2r_k\frac{(\x - \tl_k)^T\tn_k}{\norm{\x-\tl_k}}   + (1 - r_k)(\mu_k + 1)\cos(\beta_{ik} - \alpha_{ik})^{\mu_k}\right),
\label{eq:reflectedChannel}
\end{align}
where $\mi$ is the Lambertian order for the $i$th LED transmitter, $\Ar$ is the area of the PD at the VLC receiver, $\alpha_{ik}$ and $\beta_{ik}$ denote, respectively, the incidence angle for the direct path between the $i$th LED transmitter and the reflective point $\tl_k$ at the $k$th reflecting surface, and the irradiance angle for the direct path between the
reflective point $\tl_k$ at the $k$th reflecting surface and the VLC receiver, $\mu_k $ represents the directivity of reflection, and $r_k \in [0, \,1]$  denotes the proportion of diffuse component by the IRS. We define the total channel gain due to the LOS component and the reflections from the $\NR$ IRS elements from the $i$th LED to the VLC receiver as
\begin{align}
     h_i(\x,\tn_k) \triangleq H^{{\rm{LOS}}}_{i}(\x) + \sum_{k=1}^{\NR} \int_{S_{k}} dH_{i,k}^{{\rm{ref}}} (\x,\tl_k,\tn_k) .
     \label{eq:totalChannel}
\end{align}

In practical implementations, the orientation vectors of the IRS elements cannot be adjusted perfectly and 
deviate from the true (desired) orientations due to device imperfections and noise (inaccuracy) in gyroscope readings. 
Consequently, our objective is to investigate the effects of these mismatched orientation vectors of IRS elements on the accuracy of visible light positioning based on power measurements.

\section{Estimation under Mismatched Orientations: MML and MCRB}\label{sec:Est}

Let $\tn_k^m$ for $k=1,\ldots,\NR$ denote the mismatched orientation vectors of the IRS elements and let $\tn_k$ for $k=1,\ldots,\NR$ represent the true orientation vectors. In particular, the visible light system aims to set the orientations vectors of the IRS elements to $\tn_k^m$ for $k=1,\ldots,\NR$. However, due to imperfections, their actual values become $\tn_k$ for $k=1,\ldots,\NR$, which are unknown to the visible light system. 
Since the VLC receiver only has the knowledge of the mismatched orientations vectors (does not know the true orientation vectors) in this setting, the likelihood function for the position of the VLC receiver is expressed, based on \eqref{eq:powMeas}, as 
\begin{align}\label{eq:mismatchlikelihhod}
\tilde{p}(\PRX\,|\,\x)=
\left(\prod_{i=1}^\NL \frac{1}{ \sqrt{2\pi\sigma_i^2}} \right)e^{-\sum_{i=1}^{\NL}\frac{\left(\PRXi-P_{{\rm{TX}},i} H^{{\rm{LOS}}}_{i}(\x) -P_{{\rm{TX}},i} \sum_{k=1}^{\NR} \int_{S_{k}} dH_{i,k}^{{\rm{ref}}} (\x,\tl_k,\tn_k^m) \right)^2}{2\sigma_i^2}}
\end{align} 
where $\PRX$ represents a vector consisting of the power 
measurements; i.e., $\PRX=[ \PRXone\cdots\PRXlast ]^T$. The expression in \eqref{eq:mismatchlikelihhod} can be referred to as the \textit{misspecified parametric PDF for $\x$}, given that the reflected power measurements are inaccurately specified due the mismatched orientations \cite{iddrisu2023visible}.

For the purpose of comparison, we also define the true model, incorporating the true orientation vectors of the IRS elements, $\tn_k$ for $k=1,\ldots,\NR$. Hence, the true PDF can be formulated as follows:
\begin{align}\label{eq:truelikelihood}
p(\PRX)=
\left(\prod_{i=1}^\NL \frac{1}{ \sqrt{2\pi\sigma_i^2}}\right) e^{-\sum_{i=1}^{\NL}\frac{\left(\PRXi-P_{{\rm{TX}},i} H^{{\rm{LOS}}}_{i}(\xbar) -P_{{\rm{TX}},i} \sum_{k=1}^{\NR}\int_{S_{k}} dH_{i,k}^{{\rm{ref}}} (\xbar,\tl_k,\tn_k)\right)^2}{2\sigma_i^2}}
\end{align}
where $\xbar$ represents the true position vector of the VLC receiver.

\subsection{MML Estimator}
Given the misspecified parametric PDF in \eqref{eq:mismatchlikelihhod}, MML estimator, defined as the parameter that maximizes this PDF, can be expressed as \cite{fortunati2017,iddrisu2023visible}:
\begin{align}
\label{eq:misEst}
\xhatmml(\PRX) = 
\underset{\x\in\mtL}{\operatorname{arg\,max}}\;\tilde{p}(\PRX\,|\,\x) 
\end{align}
where $\mtL$ represents the set of possible positions for the VLC receiver. Based on \eqref{eq:mismatchlikelihhod} and \eqref{eq:misEst}, the MML estimator derived from the received signal power measurements $\PRX$ in \eqref{eq:powMeas} is given by
\begin{align}
\xhatmml(\PRX) = 
\underset{\x\in\mtL}{\operatorname{arg\, min}}\;\sum_{i=1}^{\NL}\frac{\left(\PRXi-P_{{\rm{TX}},i} H^{{\rm{LOS}}}_{i}(\x) -P_{{\rm{TX}},i} \sum_{k=1}^{\NR} \int_{S_{k}} dH_{i,k}^{{\rm{ref}}} (\x,\tl_k\,\tn_k)\right)^2}{\sigma_i^2}\,\cdot
\end{align}
Under appropriate regularity conditions, it is established that the MML estimator, $\xhatmml(\PRX)$, is asymptotically unbiased, and its error covariance matrix equals the $\text{MCRB}(\x_0)$, which is derived in the subsequent section \cite{fortunati2017}.

\subsection{MCRB}

To obtain the theoretical limit on the accuracy of position estimation using the MML estimator, we first define the  pseudo-true parameter \cite{fortunati2017}, which is the parameter that minimizes the Kullback-Leibler (KL) divergence between the misspecified parametric PDF $\tilde{p}(\PRX\,|\,\x)$ in \eqref{eq:mismatchlikelihhod} and the true PDF ${p}(\PRX)$ in \eqref{eq:truelikelihood}. Namely, the pseudo-true parameter is given by
\begin{equation}
\label{eq:pseudoTrue}
\x_0 = 
\underset{\x\in\mathbb{R}^3}{\operatorname{arg\,min}}  \,D(p(\PRX) \|\, \tilde{p}(\PRX\,|\,\x))
\end{equation}
where $D(p(\PRX) \|\, \tilde{p}(\PRX\,|\,\x))$ represents the KL divergence between the PDFs ${p}(\PRX)$ and $\tilde{p}(\PRX\,|\,\x)$.
The objective function in \eqref{eq:pseudoTrue} can be expressed based on the definition of the KL divergence as follows:
\begin{align}\nonumber
D(p(\PRX) \| \, \tilde{p}(\PRX\,|\,\x)) 
&= \int \log \left( { \frac{p(\PRX)}{\tilde{p}(\PRX\,|\,\x)} } \right)p(\PRX)d\PRX\\
&={\mathbb{E}}_p \bigg\{\log \left( { \frac{p(\PRX)}{\tilde{p}(\PRX\,|\,\x)} }\right) \bigg\}\cdot
\label{eq:ExpLogRatio}
\end{align}

We let $\xhat(\PRX)$ represent a misspecified-unbiased (MS-unbiased) estimator of the true position, ${\boldsymbol{\bar{x}}}$. That is, the expected value of the estimator $\xhat(\PRX)$ under the true model is equal to $\x_0$ \cite{Cuneyd}. The MCRB is a lower bound on the covariance matrix of any MS-unbiased estimator $\xhat(\PRX)$ of ${\boldsymbol{\bar{x}}}$ and is expressed as \cite{fortunati2017,FORTUNATI2018197}:
\begin{align}
\mathbb{E}_p\big\{(\xhat(\PRX) - \x_0)(\xhat(\PRX)- \x_0 )^T\big\} \succeq \text{MCRB}(\x_0) 
\end{align}
where ${\mathbb{E}}_p\{\cdot\}$  represents the expectation under the true model $p(\PRX)$ in \eqref{eq:truelikelihood}, and 
\begin{equation}
\label{eq:mcrb}
\text{MCRB}(\x_0) \triangleq \mathbf{A}_{\x_0}^{-1}\mathbf{B}_{\x_0}\mathbf{A}_{\x_0}^{-1}.
\end{equation}

The elements corresponding to the $(m,n)$th entries of matrices $\mathbf{A}_{\x_0}$ and $\mathbf{B}_{\x_0}$ in \eqref{eq:mcrb} are computed as follows:
\begin{align} 
\label{eq:matrixA}
\big[\mathbf{A}_{\x_0}\big]_{mn}&=
{\mathbb{E}}_{p}\left\{\frac{\partial^{2}\log \tilde{p}(\PRX\,|\,\x)}{\partial \x(m)\partial \x(n)}\bigg{\vert}_{\x = \x_0}\right\}, 
\\
\label{eq:matrixB}
\big[\mathbf{B}_{\x_0}\big]_{mn}&=
{\mathbb{E}}_{p} \left\{\frac{\partial \log \tilde{p}(\PRX\,|\,\x)}{\partial \x(m)}  
\frac{\partial \log \tilde{p}(\PRX\,|\,\x)}{\partial \x(n)}\bigg{\vert}_{\x =\x_0}
\right\}   
\end{align} 
for $1 \leq m,n \leq 3$, where $\x(n)$ represents the $n$th element of $\x$. Given that the value of the pseudo-true parameter typically is not of interest, the MCRB is utilized to establish a lower bound (LB) for any MS-unbiased estimator with respect to the true parameter value \cite{fortunati2017}. Specifically, the error covariance matrix for any MS-unbiased estimator $\xhat(\PRX)$ is lower bounded as follows \cite{iddrisu2023visible}:
\begin{align}
{\mathbb{E}}_p\,\big\{(\xhat(\PRX) -\xbar)( \xhat(\PRX)-\xbar )^T\big\} \, \succeq  \, \text{LB}(\x_0)
\end{align}
where 
\begin{equation} 
\label{eq:lowerBound}
\text{LB}(\x_0) \triangleq \text{MCRB}(\x_0) + (\xbar - \x_0)(\xbar- \x_0 )^T.
\end{equation}
The last term in \eqref{eq:lowerBound} represents a bias component that is independent of the SNR. Consequently, as the SNR goes to infinity, the MCRB tends toward zero, and the bias component closely approaches the error covariance matrix of the MS-unbiased estimator.

To obtain the MCRB, and consequently the LB in \eqref{eq:lowerBound}, the following derivations are carried out.

\subsubsection{Derivation of the Pseudo-True Parameter}
We first obtain the pseudo-true parameter $\x_0$, as  defined in \eqref{eq:pseudoTrue}, as it is essential for computing the MCRB associated with estimating the position of the VLC receiver in the presence of mismatched orientations. This involves finding the value of $\x$ that minimizes the KL divergence between the PDFs $\tilde{p}(\PRX\,|\,\x)$ in \eqref{eq:mismatchlikelihhod} and $p(\PRX)$ in \eqref{eq:truelikelihood}. Utilizing equations \eqref{eq:mismatchlikelihhod} and \eqref{eq:truelikelihood}, the expression in \eqref{eq:ExpLogRatio} can be computed as follows:
\begin{align}\nonumber
D(p(\PRX) \| \, \tilde{p}(\PRX\,|\,\x))
&= {\mathbb{E}}_p \Bigg\{ \sum_{i=1}^{\NL}\frac{\left(\PRXi-P_{{\rm{TX}},i} H^{{\rm{LOS}}}_{i}(\xbar) -P_{{\rm{TX}},i} \sum_{k=1}^{\NR}\int_{S_{k}} dH_{i,k}^{{\rm{ref}}} (\xbar,\tl_k,\tn_k)\right)^2}{2\sigma_i^2}
\\\nonumber&-\sum_{i=1}^{\NL}\frac{\left(\PRXi-P_{{\rm{TX}},i} H^{{\rm{LOS}}}_{i}(\x) -P_{{\rm{TX}},i} \sum_{k=1}^{\NR} \int_{S_{k}} dH_{i,k}^{{\rm{ref}}} (\x,\tl_k\,\tn_k^m \right)^2}{2\sigma_i^2}
    \Bigg\}
\\&=\sum_{i=1}^{\NL}\frac{(\mathbb{E}_{p}(\PRXi) - \mathbb{E}_{\tilde{p}}(\PRXi))^2}{2\sigma_i^2} 
\label{eq:Objsol}
\end{align}
where $\mathbb{E}_{p}(\PRXi) = P_{{\rm{TX}},i} H^{{\rm{LOS}}}_{i}(\xbar) + P_{{\rm{TX}},i} \sum_{k=1}^{\NR} \int_{S_{k}} dH_{i,k}^{{\rm{ref}}} (\xbar,\tl_k,\tn_k)$ and $\mathbb{E}_{\tilde{p}}(\PRXi) = P_{{\rm{TX}},i} H^{{\rm{LOS}}}_{i}(\x)+P_{{\rm{TX}},i} \sum_{k=1}^{\NR} \int_{S_{k}} dH_{i,k}^{{\rm{ref}}} (\x,\tl_k\,\tn_k^m)$. Based on \eqref{eq:ExpLogRatio} and \eqref{eq:Objsol}, we can find the value of  $\x$ that minimizes the Kullback-Leibler (KL) divergence as 
\begin{equation}
\label{pseudoTrueParameter}
\x_0 =
\underset{\x\in\mathbb{R}^3}{\operatorname{arg\,min}}\sum_{i=1}^{\NL}\frac{(\mathbb{E}_{p}(\PRXi) - \mathbb{E}_{\tilde{p}}(\PRXi))^2)}{\sigma_i^2} \,\cdot
\end{equation}

\subsubsection{MCRB Derivation}

After obtaining the pseudo-true parameter, the matrices $\mathbf{A}_{\x_0}$ in \eqref{eq:matrixA} and $\mathbf{B}_{\x_0}$ in \eqref{eq:matrixB} are computed to calculate the MCRB outlined in \eqref{eq:mcrb}. The second-order derivatives in \eqref{eq:matrixA} can be obtained from \eqref{eq:mismatchlikelihhod} as follows:
\begin{align}\nonumber
\frac{\partial^{2} \log \tilde{p}(\PRX\,|\,\x)}{\partial \x(n)\partial \x(m)} &= 
\sum_{i=1}^{\NL}\frac{(\PRXi\,\PTXi)}{\sigma_i^2} \frac{\partial^{2}  h_i(\x, \tn_k^m)}{\partial \x(n)\partial \x(n)}
-
\sum_{i=1}^{\NL}\frac{(\PTXi)^2}{\sigma_i^2} \frac{\partial   h_i(\x, \tn_k^m) }{\partial \x(m)} \frac{\partial  h_i(\x, \tn_k^m) }{\partial \x(n)}\\
\label{eq:SecDerA}
&-\sum_{i=1}^{\NL}\frac{(\PTXi)^2}{\sigma_i^2}  h_i(\x, \tn_k^m)\frac{\partial^{2}  h_i(\x, \tn_k^m)}{\partial \x(n)\partial \x(m)}
\end{align}
where 
$     h_i(\x, \tn_k^m) = H^{{\rm{LOS}}}_{i}(\x) + \sum_{k=1}^{\NR} \int_{S_{k}} dH_{i,k}^{{\rm{ref}}} (\x,\tl_k,\tn_k^m)$. Then, by evaluating \eqref{eq:SecDerA} at $\x=\x_0$, the elements of matrix $\mathbf{A}_{\x_0}$ in \eqref{eq:matrixA} can be derived as 
\begin{align} \nonumber
\big[\mathbf{A}_{\x_0}\big]_{mn}&=
\sum_{i=1}^{\NL}\frac{{\mathbb{E}}_{p}\{\PRXi\}\,\PTXi\,  }{\sigma_i^2} \frac{\partial^{2} h_i(\x_0, \tn_k^m)}{\partial \x(n)\partial \x(m)}
-
\sum_{i=1}^{\NL}\frac{(\PTXi\,)^2}{\sigma_i^2} \frac{\partial  h_i(\x_0, \tn_k^m) }{\partial \x(m)} \frac{\partial  h_i(\x_0, \tn_k^m) }{\partial \x(n)}\\
\label{eq:matrixAexplicit}
&-\sum_{i=1}^{\NL}\frac{(\PTXi\,)^2}{\sigma_i^2} h_i(\x_0, \tn_k^m)\frac{\partial^{2} h_i(\x_0, \tn_k^m)}{\partial \x(n)\partial \x(m)}\,\cdot
\end{align}

The first-order derivatives in \eqref{eq:matrixB} can be obtained from \eqref{eq:mismatchlikelihhod} as follows: 
\begin{align}\label{eq:firstDerMatB}
\frac{\partial \log \tilde p(\PRX\,|\,\x) }{\partial \x(m)}= \sum_{i=1}^{\NL} \frac{\PTXi(\PRXi-P_{TX,i}h_i(\x, \tn_k^m)}{\sigma_i^2} \frac{\partial h_i(\x, \tn_k^m)}{\partial \x(m)}\,\cdot
\end{align}
Also, the elements of matrix $\mathbf{B}_{\x_0}$ in \eqref{eq:matrixB} are obtained utilizing \eqref{eq:firstDerMatB} as follows:
\begin{align}\nonumber
\big[\mathbf{B}_{\x_0}\big]_{mn}&=
{\mathbb{E}}_{p} \Bigg\{\sum_{i=1}^{\NL} \PTXi\, \frac{(\PRXi-\PTXi\,  h_i(\x_0, \tn_k^m))}{\sigma_i^2} \frac{\partial h_i(\x_0, \tn_k^m)}{\partial \x(m)}
\\&\times
\sum_{j=1}^{\NL} \PTXj \, \frac{(\PRXj-\PTXj \,  h_j(\x_0,\tn_k^m))}{\sigma_j^2} \frac{\partial h_j(\x_0,\tn_k^m)}{\partial \x(n)}
\Bigg\}
\\\nonumber
&=\sum_{i=1}^{\NL} \left(\frac{\PTXi\, }{\sigma_i^2} \right)^2
{\mathbb{E}}_{p}\{(\PRXi-\PTXi\,  h_i(\x_0, \tn_k^m))^2\}\frac{\partial h_i(\x_0, \tn_k^m)}{\partial \x(m)}\frac{\partial h_i(\x_0, \tn_k^m)}{\partial \x(n)} \\ \nonumber
&+\sum_{i=1}^{\NL}\sum_{j=1
,j \neq i}^{\NL}\frac{\PTXi\,\PTXj}{\sigma_i^2\sigma_j^2} (\PTXi\, h_i(\xbar,\tn_k) -\PTXi\,  h_i(\x_0, \tn_k^m) )
\\\label{eq:matrixBexplicit}
&\times
(\PTXj h_j(\xbar,\tn_k) -\PTXj\,  h_j(\x_0,\tn_k^m) ) \frac{\partial h_i(\x_0, \tn_k^m)}{\partial \x(m)}\frac{\partial h_j(\x_0,\tn_k^m)}{\partial \x(n)}
\end{align}
where ${\mathbb{E}}_{p}\{(\PRXi-\PTXi\,  h_i(\x_0,\tn_k^m))^2\} = \sigma_i^2 + (\PTXi\, h_i(\xbar,\tn_k^m) -\PTXi\,  h_i(\x_0,\tn_k^m) )^2$.

The first and second order derivatives of the channel coefficient $h_i(\cdot)$, as indicated in \eqref{eq:matrixAexplicit} and \eqref{eq:matrixBexplicit}, can be explicitly obtained with respect to the system parameters, as detailed in \textbf{Appendix~\ref{app:ChanCoefDer}}.

\textbf{Remark:} Based on the derived MCRB and LB expressions, we can evaluate the impact of mismatched orientations of IRS elements; namely, the effects of imperfections in IRS orientation adjustment.
For comparison purposes, we can examine the maximum likelihood (ML) estimator and the corresponding Cramér-Rao bound (CRB), as derived without orientation mismatches in \cite[Sec.~II]{kokdogan2024intelligent}. Since the theoretical limits are achieved by the MML and ML estimators at high SNRs, they serve as benchmarks for assessing localization performance and for designing system parameters under high SNR conditions.

\section{Numerical Results}\label{sec:Nume}

\begin{figure}
\includegraphics[width=0.9\linewidth]{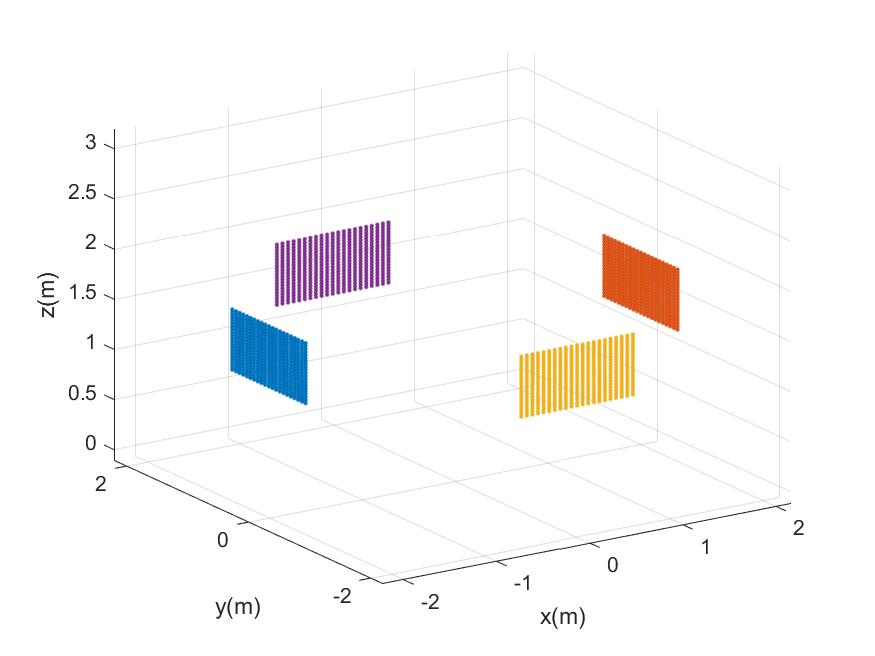}
\centering
\caption{A depiction of the IRS elements on the walls of the room.}
 \label{fig:IRSelements}
\end{figure}

In this section, we conduct simulations to investigate the effects of mismatched orientations of IRS elements on localization accuracy by evaluating the proposed estimator and the theoretical bound. As in \cite{kokdogan2024intelligent}, we consider a room with dimensions of $4\times 4\times 3$ meters (width, depth, and height, respectively) and $4$ LEDs ($\NL = 4$) placed at the following coordinates in meters: $(-1, 1, 3),(1, 1, 3),(1, -1, 3),(-1, -1, 3)$. These LED positions are selected to ensure a symmetrical coverage of the room, with the center of the room floor being represented by the coordinates $(0, 0, 0)$. The orientations of the LEDs are defined as $\Ni = [0, 0,-1]^T$ for $i = 1,\ldots, \NL$, indicating that they all point downwards. The transmit powers of the LEDs, $\PTXi$, are all set to $5\,$W  and the area of the PD is taken as $\Ar = 1\,{\rm{cm}}^2$. The orientation vector of the receiver is defined as $\nR = [0, 0, 1]^T$, indicating that it points upwards. Furthermore, we assume that the noise variances are equal, denoted as $\sigma_i^2 = \sigma^2$ for all $i$ (see \eqref{eq:powMeas}). As depicted in Fig.~\ref{fig:IRSelements}, each of the four walls has $\NR/4$ IRS units arranged in a $\sqrt{\NR/4} \times \sqrt{\NR/4}$ grid. In the simulations, $\NR$ is set to $1764$, resulting in $21\times21$ IRS elements per wall \cite{kokdogan2024intelligent}. Each IRS element has a rectangular shape with a width $w_u$ of $4\,$cm and a height $h_u$ of $2\,$cm, giving each IRS element an area $S_k$ of $8\,$cm². The IRS units are spaced $w_d = 2\,$cm apart horizontally and $h_u = 1 \,$cm apart vertically to prevent inter-element blockage, as detailed in Fig.~3 of \cite{kokdogan2024intelligent}. The IRS units are positioned to center both horizontally and vertically on the wall, aligning the center of the two-dimensional IRS array with the center of the wall. In the simulations, the fraction of diffuse component, $r_k$, is set to $0$ and the directivity of reflection, $\mu_k$, is set to $5$, and the reflectance coefficient of the IRS elements is set to $\rho_k = 0.95\,\, \forall k$. In addition, the mismatched orientations (the imperfect orientation information at the VLC receiver) are set for all $k$'s as 
$\tn_k^m = [0~ 1 ~0]^T$ $(\theta_1 = \pi/2,\,
\phi_1 =  0)$, $\tn_k^m = [0\,-1~ 0]^T$ $(\theta_2 =  \pi/2,\,\phi_2 =  \pi)$, $\tn_k^m = [1~ 0 ~0]^T$ $(\theta_3 = \pi/2,\,\phi_3 =  \pi/2)$ and $\tn_k^m = [-1~ 0~ 0]^T$ $(\theta_4 = \pi/2,\,\phi_4 = -\pi/2)$ for the IRS elements located on the wall at $y = -2$, $y = 2$, $x = -2$, and $ x = 2$, respectively. On the other hand, the true orientations are obtained as $\tn_k = [\cos(\phi^t_i)\sin(\theta^t_i) ~ \sin(\phi^t_i)\sin(\theta^t_i) ~ \cos(\theta^t_i)]^T$ with $\theta_i^t = \theta_i  + {\mathcal{U}}(-\rm{k},\rm{k})$, $\phi_i^t=  \phi_i + {\mathcal{U}}(-\rm{k},\rm{k})$ for $i \in \{1,2,3,4\}$, where ${\mathcal{U}}(-\rm{k},\rm{k})$ denotes a uniform random variable over the interval $[-\rm{k},\rm{k}]$. This setting corresponds to a scenario where the visible light system aims to set the orientation vectors of the IRS elements to be perpendicular to the walls; however, due to imperfections, the true orientations are not perfectly perpendicular as specified above.

\begin{figure}
\includegraphics[width=1\linewidth]{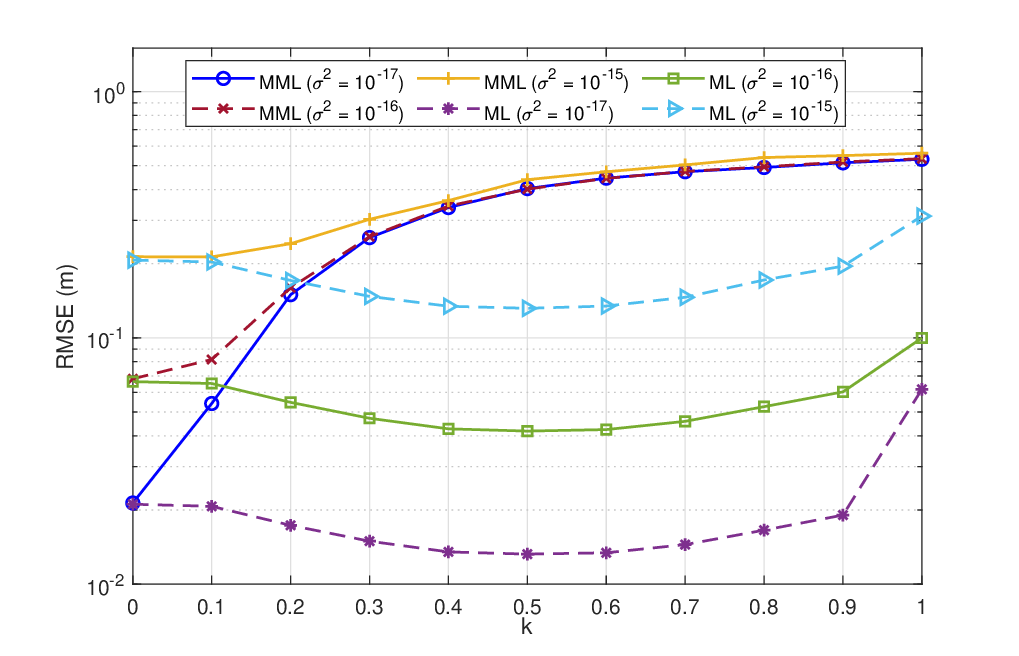}
\centering
\caption{Performance of the MML estimator versus k for $\sigma^2 \in \{10^{-15},10^{-16} ,10^{-17}\}$ in NLOS scenario when the VLC receiver is positioned at (0.5,\, 0.5,\,  0.85) meters.} 
 \label{fig:MMLvsk}
\end{figure}

To evaluate the performance of the MML estimator and to compare it against the ML estimator that is implemented assuming perfectly set (i.e., perpendicular) IRS orientations, the VLC receiver is placed at $\x = (0.5,\, 0.5,\,  0.85)\,$m. In Fig.~ \ref{fig:MMLvsk}, we plot the root mean-squared error (RMSE) performance of the MML estimator and the ML estimator versus $\rm{k}$ for three different noise levels: $10^{-15}$, $10^{-16}$, and $10^{-17}$. For each $\rm{k}$ value, the simulation is run 500 times to calculate the RMSE. It can be observed that the RMSE values of the MML estimator for all noise levels tend to rise as $\rm{k}$ increases. This is due to the fact that the mismatch between the true IRS element orientations and the assumed orientations increases with  $\rm{k}$, consequently leading to degraded performance of the MML estimator. Additionally, the MML estimator exhibits its worst RMSE performance at $\sigma^2 = 10^{-15}$ compared to its performance at $\sigma^2 = 10^{-16}$ and $\sigma^2 =10^{-17}$, due to the higher noise content in the received power measurements. The RMSE values of the MML estimator for all three noise levels when $\rm{k} = 0$ coincides with the true model as there is no mismatch between the actual and assumed orientation vectors of the IRS elements. It is observed that as $\rm{k}$ increases, the RMSE performance does not improve significantly with decreasing noise variance as the mismatch between the assumed and true orientations of the IRS elements becomes the primary source of error. Moreover, it is worth noting that the RMSE performance of the ML estimator does not follow a monotone pattern with respect to $\rm{k}$. The optimal IRS orientations depend on the actual position of the VLC receiver; hence, RMSE values of the ML estimator vary, being high for certain $\rm{k}$ values and low for others. On average, the optimal orientations of the IRS elements are achieved at around $\rm{k} = 0.5$ when the VLC receiver is located at $(0.5,\, 0.5,\,  0.85)\,$m. The RMSE values tend to rise for $\rm{k}$ values greater or less than $0.5$.
This outcome is consistent with the findings in \cite{kokdogan2024intelligent}, which developed an algorithm for optimally adjusting the IRS orientations and demonstrated that IRS perpendicularity is not optimal for the position $(0.5,\, 0.5,\,  0.85)$. Furthermore, unlike the MML estimator, with increasing  $\rm{k}$ values, the RMSE performance of the ML estimator improves significantly with decreasing noise levels as it is not affected by the orientation mismatches.

\begin{figure}
\includegraphics[width=1\linewidth]{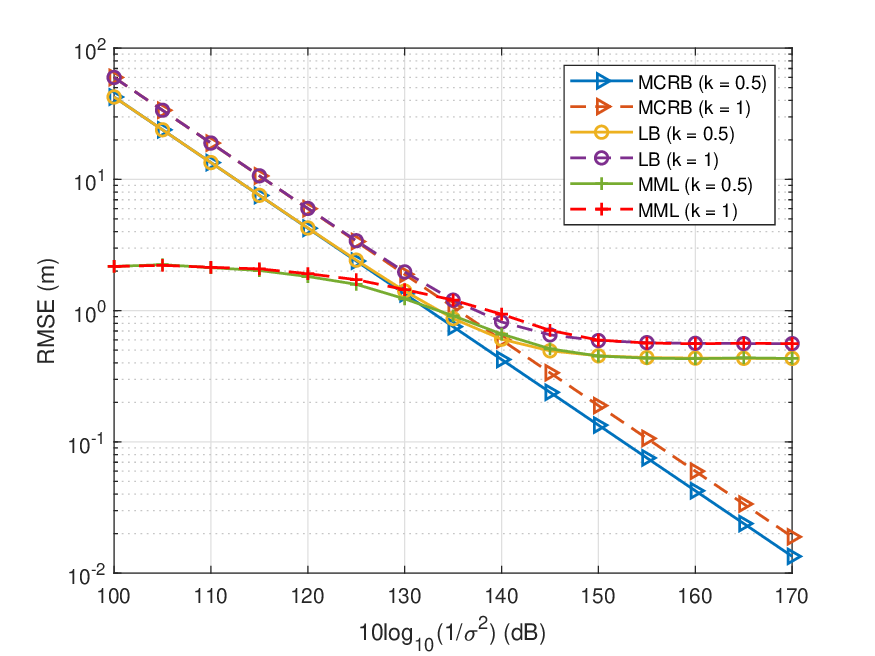}
\centering
\caption{RMSE performance of the MML estimator and the theoretical limits versus $10\log_{10}(1/\sigma^2)$ for $ \rm{k} \in \{0.5, 1\}$  in NLOS scenario  when the VLC receiver is positioned at (0.5,\, 0.5,\,  0.85) meters.}
 \label{fig:mcrbmlvsnoise}
\end{figure}

In Fig.~ \ref{fig:mcrbmlvsnoise}, we plot the RMSE performance of the MML estimator and the theoretical bounds versus $10\log_{10}(1/\sigma^2)$ for different $\rm{k}$ values of $0.5$ and $1$. The orientations of the IRS elements are generated using MATLAB with a seed value of $1$. It can be observed that for both values of $\rm{k}$, the LBs plateau at low noise variances and the MCRBs approach zero, as anticipated. Additionally, it is worth noting that the MCRB and LB values at $\rm{k} = 0.5$ are lower than those at $\rm{k} = 1$. This is due the greater mismatch of orientation vectors at $\rm{k} = 1$ compared to $\rm{k} = 0.5$. It can also be observed that the RMSE of the MML estimator tends to converge to the LB at low noise variances for both values of $\rm{k}$. This occurs because, at low noise variances, estimation errors are mainly attributed to orientation mismatches and remain consistent even as noise variance decreases. However, at high noise variances, the RMSE of the MML estimator is observed to be lower than the established bounds. This can be attributed to the fact that the location of the VLC receiver is restricted to a room with dimensions $4 \times 4\times 3$ meters, and the MML estimator performs search within this region. The calculations made within this region relying on random estimates for the position of the VLC receiver yield an RMSE of approximately $2.1$ meters, closely aligning with the estimation error of the estimators under high noise level conditions.

\begin{figure}
\includegraphics[width=1\linewidth]{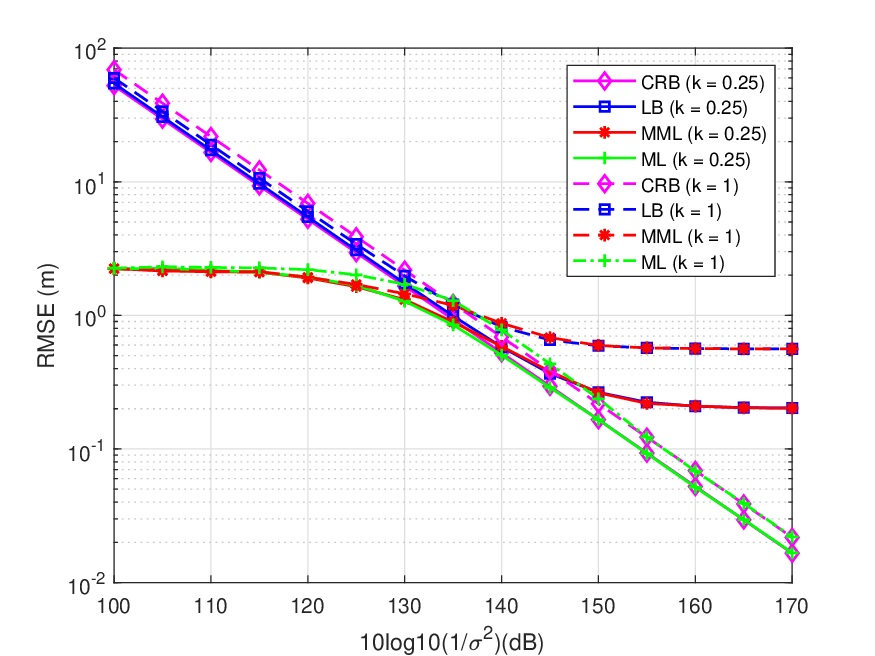}
\centering
\caption{RMSE performance of the MML estimator and the theoretical limits versus $10\log_{10}(1/\sigma^2)$ for $ \rm{k} \in \{0.25, 1\}$  in NLOS scenario and the VLC receiver is positioned at (0.5,\, 0.5,\,  0.85) meters.}
 \label{fig:MMLMLvsnoise}
\end{figure}

In Fig.~ \ref{fig:MMLMLvsnoise}, we plot the RMSE performance of the MML estimator and ML estimator, and their corresponding theoretical bounds versus $10\log_{10}(1/\sigma^2)$ for $\rm{k} \in \{0.25, 1\} $, with the orientations of the IRS elements generated using MATLAB and a seed of $1$. As anticipated, the LBs plateau at low noise variances, and the MCRBs and CRBs approach zero for both values of $\rm{k}$. It can also be observed that the RMSE values of the MML estimator and ML estimator for both values of $\rm{k}$ tend to converge to the LB and the CRB, respectively, at low noise variances. The RMSE values of MML estimator when $\rm{k} = 0.25$ are lower than those for $\rm{k} = 1$ due to the increased orientation mismatch. However, for the ML estimator, the RMSE values are lower when $\rm{k} = 0.25$ than when $\rm{k} = 1$ due to fact that the generated orientations when $\rm{k} = 0.25$ are closer to optimal for the VLC receiver location than when $\rm{k} = 1$. Similar to what has been observed in Fig.~ \ref{fig:mcrbmlvsnoise}, at high noise variances, the RMSEs of the MML estimator and ML estimator are observed to be lower than their established bounds due to the restriction on the search region for the estimators, as previously discussed. Additionally, it can be observed at very low noise variances that the ML estimator outperforms the MML estimator for both values of $\rm{k}$, as the orientation mismatches become the primary source of error under the conditions of low noise variances.

\FloatBarrier

\section{Concluding Remarks}\label{sec:Conc}

A position estimation problem has been formulated based on received power measurements for an IRS-assisted VLP system with a mismatch between the true and assumed orientations of IRS elements. The MML estimator, MCRB and LB have been derived to quantify the degradation in performance due to orientation mismatches. To assess the performance degradation of the MML estimator arising from orientation mismatches, a comparative analysis has been conducted against the ML estimator and the CRB under the scenario of no orientation mismatches. By comparing the MML and the LB with the ML estimator and the CRB, we have demonstrated the impact of orientation mismatches on position estimation performance. Specifically, orientation mismatches lead to a significant performance loss at high SNRs, thus hindering the achievement of very low estimation errors.

\appendix
\section{Appendices}

\subsection{First and Second Derivatives of Channel Coefficient}\label{app:ChanCoefDer}

The first-order derivatives of the channel coefficient $h_i(\x,\tn_k)$ appearing in \eqref{eq:matrixAexplicit} and \eqref{eq:matrixBexplicit} with respect to the position parameter $\x$ can be calculated using \eqref{eq:totalChannel} as follows:
\begin{align}
     \frac{\partial h_i(\x, \tn_k)}{\partial \x(m)}= \frac{\partial H^{\rm{LOS}}_{i}(\x)}{\partial \x(m)}  + \frac{\sum_{k=1}^{\NR} \int_{S_{k}} \partial dH_{i,k}^{\rm{ref}} (\x,\tl_k,\tn_k)}{\partial \x(m)}
\label{eq:firstDerivative}
\end{align}
where
\begin{align}
\frac{\partial H^{\rm{LOS}}_{i}(\x)}{\partial \x(m)}=\frac{(\mi+1) \Ar}{2\pi}
\left(\frac{1}{v(\x)}\frac{\partial u(\x) }{\partial \x(n)} 
- \frac{u(\x)}{v^2(\x)}\frac{\partial v(\x) }{\partial \x(n)}
\right)
\end{align}
with
\begin{align}\label{eq:ulr}
u(\x) &= \left[(\x - \lt_i)^{T} \Ni \right]^{\mi} (\lt_i-\x)^{T}\nR,
\\\label{eq:vlr}
v(\x) &= \norm{\x - \lt_i}^{\mi+3},
\\\label{eq:ulr_der}
\frac{\partial u(\x) }{\partial \x(n)} &=\mi((\lt_i-\x)^{T}\nR)
\left[(\x - \lt_i)^{T} \Ni \right]^{\mi - 1}\Ni(n)\, -
\left[(\x - \lt_i)^{T} \Ni \right]^{\mi}\nR(n),
\\\label{eq:vlr_der}
\frac{\partial v(\x) }{\partial \x(n)}&= (\mi+3)(\x(n)- \lt_i(n))\norm{\x - \lt_i}^{\mi+1},
\end{align}
and 
\begin{align}\nonumber
&\frac{\partial  dH_{i,k}^{\rm{ref}}(\x,\tl_k\tn_k)}{\partial \x(m)}=   \frac{(\mi+1) \left[(\tl_k - \lt_i)^{T} \Ni \right]^{\mi} ((\lt_i-\tl_k)^{T} \tn_k)\rho_k dS_k}{4\pi^2\norm{\lt_i - \tl_k}^{\mi+3}}
 \\&\nonumber \times \Bigg\{
2r_k\Bigg [ \frac{\tn_k(m)(\tl_k - \x)^T\nR-\nR(m)(\x - \tl_k)^T\tn_k}{\norm{\x -\tl_k}^4} - \frac{4(\x(m)- \tl_k(m))(\x- \tl_k)^T\tn_k(\tl_k- \x)^T\nR}{\norm{\x -\tl_k}^6} \Bigg ] +\\&\nonumber
(1 - r_k)(\mu_k + 1)\bigg[(\cos(\beta_{ik} - \alpha_{ik}))^{\mu_k}(-\nR(m)\norm{\x-\tl_k}^{-3}-3(\tl_k-\x)^T\nR\norm{\x-\tl_k}^{-5}(\x(m)- \tl_k(m)))+ \\&\mu_k\cos(\beta_{ik} - \alpha_{ik})^{\mu_k-1} \frac{(\tl_k - \x)^{T} \nR }{\norm{\x -\tl_k}^3} \frac{\partial \cos(\beta_{ik} - \alpha_{ik})}{\partial \x(m)} \bigg]
\Bigg\}
\end{align}
with 
\begin{multline}
\cos(\beta_{ik} - \alpha_{ik}) = \cos (\alpha_{ik})\bigg(((\x - \tl_k)^{T} \tn_k)\norm{\x - \tl_k}^{-1} \bigg) + \sin (\alpha_{ik}) \bigg (\norm{(\x - \tl_k) \times \tn_k} \norm{\x - \tl_k}^{-1}\bigg ),
\end{multline}
\begin{multline}
\frac{\partial \cos(\beta_{ik} - \alpha_{ik})}{\partial \x(m)} = \cos (\alpha_{ik})\bigg(\tn_k(m) \norm{\x - \tl_k}^{-1} -(\x - \tl_k)^{T} \tn_k\|\x - \tl_k\|^{-3}(\x(m) - \tl_k(m)) \bigg)\\
+ \sin (\alpha_{ik}) \bigg (\norm{(\x - \tl_k) \times \tn_k}^{-1} \norm{\x - \tl_k}^{-1}g(m)
- \norm{\x - \tl_k}^{-3}(\x(m) - \tl_k(m)) \norm{(\x - \tl_k) \times \tn_k}\bigg ),
\end{multline}
\begin{align}
\cos (\alpha_{ik}) &= ((\lt_i-\tl_k)^{T} \tn_k)\norm{\lt_i - \tl_k}^{-1},
\\
\sin (\alpha_{ik}) &=\norm {((\lt_i-\tl_k) \times \tn_k)}\norm{\lt_i - \tl_k}^{-1},
\end{align} 
\begin{align}\nonumber
    g(m) =\tn_k(z(m+1))) [(\x(m) - \tl_k(m)) \tn_k(z(m+1))) - (\x(z(m+1))) - \tl_k(z(m+1))))\tn_k(m)] \\ - \tn_k(z(m+2)) [(\x(z(m+2)) - \tl_k(z(m+2)))\tn_k(m) - (\x(m) - \tl_k(m))\tn_k(z(m+2))],
\end{align} 
\begin{equation}
z(m) = \begin{cases}
m, & \text{if } m \leq 3 \\
m - 3, & \text{otherwise}
\end{cases}.
\end{equation}

In addition, the second-order partial derivatives of the channel coefficient with respect to the position parameter can be derived from \eqref{eq:firstDerivative} as follows:
\begin{align}
     \frac{\partial^2 h_i(\x, \tn_k)}{\partial \x(n)\partial \x(m)}= \frac{\partial^2 H^{\rm{LOS}}_{i}(\x)}{\partial \x(n)\partial \x(m)}  + \frac{\sum_{k=1}^{\NR} \int_{S_{k}} \partial^2 dH_{i,k}^{\rm{ref}} (\x,\tl_k,\tn_k)}{\partial \x(n)\partial \x(m)}   
     \label{eq:secondDerativeChannel}
\end{align}
where $\frac{\partial^2 H^{\rm{LOS}}_{i}(\x)}{\partial \x(n)\partial \x(m)}$ is defined by 
\begin{align}\nonumber
\frac{\partial^2 H^{\rm{LOS}}_{i}(\x)}{\partial \x(n)\partial \x(m)} = &
\frac{(\mi+1) \Ar}{2\pi}
\Bigg(
\frac{1}{v(\x)}\frac{\partial^2 u(\x) }{\partial \x(m)\partial \x(n)} - \frac{1}{v^2(\x)}\frac{\partial v(\x) }{\partial \x(m)}\frac{\partial u(\x) }{\partial \x(n)} \\
&-\frac{1}{q(\x)}\frac{\partial b_n(\x) }{\partial \x(m)} +\frac{b_n(\x)}{q^2(\x)}\frac{\partial q(\x) }{\partial \x(m)}
\Bigg)
\end{align}
with $\frac{\partial^2 u(\x) }{\partial \x(m)\partial \x(n)}$, $b_n(\x)$, $q(\x)$, $\frac{\partial b_n(\x) }{\partial \x(m)}$, and $\frac{\partial q(\x) }{\partial \x(m)}$ being given as follows:
\begin{align}\nonumber
&\frac{\partial^2 u(\x) }{\partial \x(m)\partial \x(n)} = \mi(\mi-1)((\lt_i-\x)^{T}\nR)
\left[(\x - \lt_i)^{T} \Ni \right]^{\mi - 2}\Ni(n)\Ni(m)  
\\ 
&{\hspace{2.63cm}}-\mi\left[(\x - \lt_i)^{T} \Ni \right]^{\mi - 1}(\Ni(n)\nR(m) +\Ni(m)\nR(n))
\\
&b_n(\x) \triangleq (\mi+3)(\x(n)- \lt_i(n))\left[(\x - \lt_i)^{T} \Ni \right]^{\mi}((\lt_i-\x)^{T}\nR) \\
&q(\x) \triangleq  \norm{\x - \lt_i}^{\mi+5}
\\
&\frac{\partial b_n(\x) }{\partial \x(m)} =
\begin{cases}\small
(\mi+3)(\x(n)- \lt_i(n))[\mi\nT(m)((\lt_i-\x)^{T}\nR)
\left((\x - \lt_i)^{T} \Ni \right)^{\mi-1} \\\small-\nR(m)\left((\x - \lt_i)^{T} \Ni \right)^{\mi} ],
~{\hspace{7.8cm}}{\rm{if}}~m\neq n
\\\small
(\mi + 3)\left[(\x - \lt_i)^{T} \Ni \right]^{\mi}((\lt_i-\x)^{T}\nR) + (\mi+3)(\x(m)- \lt_i(m)) \\\small\times\left[\mi\nT(m)((\lt_i-\x)^{T}\nR)
\left((\x - \lt_i)^{T} \Ni \right)^{\mi - 1} -\nR(m)\left((\x - \lt_i)^{T} \Ni \right)^{\mi} \right],
~{\rm{if}}~m= n
\end{cases}\normalsize
\\\label{eq:der_q}
&\frac{\partial q(\x) }{\partial \x(m)} = (\mi+5)(\x(m)- \lt_i(m))\norm{\x - \lt_i}^{\mi+3}.
\end{align}
Also, $\frac{\partial^2  dH_{i,k}^{\rm{ref}}(\x,\tl_k\tn_k)}{\partial \x(n)\partial \x(m)}$ in \eqref{eq:secondDerativeChannel} is defined as 
\begin{align}\nonumber
\frac{\partial^2  dH_{i,k}^{\rm{ref}}(\x,\tl_k\tn_k)}{\partial \x(n)\partial \x(m)} = &c \bigg (\frac{\partial^2 f_1(\x)}{\partial\x(n) \x(m)}(f_2(\x)+f_3(\x)) +  
\frac{\partial f_1(\x)}{\partial \x(m)} (\frac{\partial f_2(\x)}{\partial \x(n)} + \frac{\partial f_2(\x)}{\partial \x(n)}) \\&+
\frac{\partial f_1(\x)}{\partial \x(n)} (\frac{\partial f_2(\x)}{\partial \x(m)} + \frac{\partial f_2(\x)}{\partial \x(m)}) + 
f_1(\x)(\frac{\partial^2 f_2(\x)}{\partial\x(n) \x(m)} + \frac{\partial^2 f_3(\x)}{\partial\x(n) \x(m)})\bigg )
\end{align}
with $c$, $f_1(\x) $, $f_2(\x)$, and $f_3(\x)$ being expressed as 
\begin{align}
c = \frac{(\mi+1) \Ar\rho_k dS_k\left[(\tl_k - \lt_i)^{T} \Ni \right]^{\mi} ((\lt_i-\tl_k)^{T} \tn_k)}{4\pi^2\norm{\lt_i - \tl_k}^{\mi+3}},
\end{align}
\begin{align}\label{eq:f1}
 f_1(\x)  =  \frac{((\tl_k-\x)^{T} \nR)}{\norm{\x - \tl_k}^{3}}, 
\end{align}
\begin{align}\label{eq:f2}
 f_2(\x) = 2r_k\frac{(\x - \tl_k)^T\tn_k}{\norm{\x-\tl_k}}, \end{align}
\begin{align}\label{eq:f3}
f_3(\x) =  (1 - r_k)(\mu_k + 1)\cos(\beta_{ik} - \alpha_{ik})^{\mu_k}.  
\end{align}

The first-order partial derivatives of $f_1(\x)$, $f_2(\x) $, and $f_3(\x)$ in \eqref{eq:f1}--\eqref{eq:f3} with respect to $\x$ are obtained as follows:
\begin{align}
\frac{\partial f_1(\x))}{\partial \x(m)} = \frac{-\nR(m)}{\norm{\x-\tl_k}^{3}}-\frac{3(\x(m)- \tl_k(m))((\tl_k-\x)^T\nR)}{\norm{\x-\tl_k}^{5}},  
\end{align}
\begin{align}
\frac{\partial f_2(\x))}{\partial \x(m)} =  2r_k\bigg(\frac{\tn_k(m)}{\norm{\x-\tl_k}}-\frac{(\x(m)- \tl_k(m))((\x -\tl_k)^T\tn_k)}{\norm{\x-\tl_k}^{3}} \bigg),
\end{align}
\begin{align}
\frac{\partial f_3(\x))}{\partial \x(m)} =   \mu_k(1 - r_k)(\mu_k + 1)\cos(\beta_{ik} - \alpha_{ik})^{\mu_k-1}\frac{\partial \cos(\beta_{ik} - \alpha_{ik})}{\partial \x(m)}.
\end{align}
Moreover, the second-order derivatives of $f_1(\x)$, $f_2(\x) $, and $f_3(\x)$ in \eqref{eq:f1}--\eqref{eq:f3} with respect to $\x$ are also obtained as follows:
\begin{align}\nonumber
\frac{\partial^2 f_1(\x))}{\partial\x(n) \x(m)} =  \frac{3\nR(m)(\x(n)- \tl_k(n))}{\norm{\x-\tl_k}^{5}}+3(\x(m)- \tl_k(m))\bigg[\frac{\nR(n)}{\norm{\x-\tl_k}^{5}} \\+\frac{5(\x(n)- \tl_k(n))((\tl_k-\x)^T\nR)}{\norm{\x-\tl_k}^{7}} \bigg],
\label{eq:f1_firstder}
\end{align}
\begin{align}
\frac{\partial^2 f_2(\x))}{\partial\x(n) \x(m)} =  -2r_k\bigg(\frac{\tn_k(m)(\x(n)-\tl_k(n)) +(\x(m)- \tl_k(m))\tn_k(n)}{\norm{\x-\tl_k}^3}\\-\frac{3(\x(n)- \tl_k(n))(\x(m)- \tl_k(m))((\x -\tl_k)^T\tn_k)}{\norm{\x-\tl_k}^{5}} \bigg) ,
\label{eq:f2_firstder}
\end{align}
\begin{align}\nonumber
\frac{\partial^2 f_3(\x))}{\partial\x(n) \x(m)} = \mu_k(1 - r_k)(\mu_k + 1)\bigg ( (\mu_k-1)\cos(\beta_{ik} - \alpha_{ik})^{\mu_k-2}\frac{\partial \cos(\beta_{ik} - \alpha_{ik})}{\partial \x(m)}\frac{\partial \cos(\beta_{ik} - \alpha_{ik})}{\partial \x(n)}\\ + \cos(\beta_{ik} - \alpha_{ik})^{\mu_k-1}\frac{\partial^2 \cos(\beta_{ik} - \alpha_{ik})}{\partial\x(n) \x(m)}\bigg ),
\label{eq:f3_firstder}
\end{align}
with
\begin{multline}
\frac{\partial^2 \cos(\beta_{ik} - \alpha_{ik})}{\partial \x(n) \x(m)} = \frac{\cos(\alpha_{ik})}{2r_k}\frac{\partial^2 f_2(\x))}{\partial \x(n) \x(m)} + \sin (\alpha_{ik})\times\begin{cases}
    A & \text{if } n = z(m+1)) \\
    B & \text{if } n = z(m+2)
\end{cases}
\label{eq:cos_secder}
\end{multline}
where the values of $A$ and $B$ in \eqref{eq:cos_secder} are as stated follows:
\begin{multline}
    A = \frac{-\tn_k(z(m+1)))\tn_k(m)}{\norm{(\x - \tl_k) \times \tn_k} \norm{\x - \tl_k}} -\frac{g(n)g(m)}{\norm{(\x - \tl_k) \times \tn_k}^{3} \norm{\x - \tl_k}}\\ -\frac{g(n)((\x(m) - \tl_k(m))) + g(m)((\x(n) - \tl_k(n)))}{\norm{(\x - \tl_k) \times \tn_k} \norm{\x - \tl_k}^{3}}\\ +\frac{3(\x(m) - \tl_k(m))(\x(n) - \tl_k(n))\norm{(\x - \tl_k) \times \tn_k}}{\norm{\x - \tl_k}^{5}},
\end{multline}
\begin{multline}
    B = \frac{-\tn_k(z(m+2))\tn_k(m)}{\norm{(\x - \tl_k) \times \tn_k} \norm{\x - \tl_k}} -\frac{g(n)g(m)}{\norm{(\x - \tl_k) \times \tn_k}^{3} \norm{\x - \tl_k}}\\ -\frac{g(n)((\x(m) - \tl_k(m))) + g(m)((\x(n) - \tl_k(n)))}{\norm{(\x - \tl_k) \times \tn_k} \norm{\x - \tl_k}^{3}}\\ +\frac{3(\x(m) - \tl_k(m))(\x(n) - \tl_k(n))\norm{(\x - \tl_k) \times \tn_k}}{\norm{\x - \tl_k}^{5}}.
    \label{eq:caseB}
\end{multline}

When $m = n$, \eqref{eq:f1_firstder}, \eqref{eq:f2_firstder}, and \eqref{eq:f3_firstder} respectively become
\begin{align}\nonumber
\frac{\partial^2 f_1(\x))}{\partial \x^2(m)} =  \frac{3\nR(m)(\x(m)- \tl_k(m))}{\norm{\x-\tl_k}^{5}}-\frac{3(((\tl_k-\x)^T\nR)- \nR(m)(\x(m)- \tl_k(m)))}{\norm{\x-\tl_k}^{5}} \\+\frac{15(\x(m)- \tl_k(m))^2((\tl_k-\x)^T\nR)}{\norm{\x-\tl_k}^{7}},   
\end{align}
\begin{align}\nonumber
\frac{\partial^2 f_2(\x))}{\partial \x^2(m)} = 2r_k\bigg(\frac{-\tn_k(m)(\x(m)-\tl_k(m)) -(\x- \tl_k)^T\tn_k -(\x(m)- \tl_k(m))\tn_k(m)}{\norm{\x-\tl_k}^3}\\+\frac{3(\x(m)- \tl_k(m))^2((\x -\tl_k)^T\tn_k)}{\norm{\x-\tl_k}^{5}} \bigg),
\end{align}
\begin{align}\nonumber
\frac{\partial^2 f_3(\x))}{\partial \x^2(m)} = \mu_k(1 - r_k)(\mu_k + 1)\bigg [ (\mu_k-1)\cos(\beta_{ik} - \alpha_{ik})^{\mu_k-2}(\frac{\partial \cos(\beta_{ik} - \alpha_{ik})}{\partial \x(m)})^2\\ + \cos(\beta_{ik} - \alpha_{ik})^{\mu_k-1}\frac{\partial^2 \cos(\beta_{ik} - \alpha_{ik})}{\partial \x^2(m)}\bigg ],
\end{align}
with 
\begin{multline}
\frac{\partial^2 \cos(\beta_{ik} - \alpha_{ik})}{\partial \x^2(m)} = \frac{\cos(\alpha_{ik})}{2r_k}\frac{\partial^2 f_2(\x))}{\partial \x^2(m)} + \sin (\alpha_{ik}) \bigg ( \frac{\tn_k^2(z(m+1)))+\tn_k^2(z(m+2))}{\norm{(\x - \tl_k) \times \tn_k} \norm{\x - \tl_k}} \\ -\frac{g^2(m)}{\norm{(\x - \tl_k) \times \tn_k}^{3} \norm{\x - \tl_k}} -\frac{g(m)(1+(\x(m) - \tl_k(m)))}{\norm{(\x - \tl_k) \times \tn_k} \norm{\x - \tl_k}^{3}}\\ -\frac{\norm{(\x - \tl_k) \times \tn_k}}{\norm{\x - \tl_k}^{3}}+\frac{3(\x(m) - \tl_k(m))^2\norm{(\x - \tl_k) \times \tn_k}}{\norm{\x - \tl_k}^{5}}
\bigg ).
\end{multline}
While the first-order derivative of the channel coefficient $h_i(\x, \tn_k)$ was provided in \cite[Eqn. (12)]{kokdogan2024intelligent}, the second-order derivatives as specified in  \eqref{eq:secondDerativeChannel}-\eqref{eq:caseB} have not been available in the existing literature.

\bibliographystyle{IEEEtran}
\bibliography{Ref_proc_IEEE}

\end{document}